\documentclass[12pt]{article}
\usepackage{epsfig,a4}
 \advance\oddsidemargin by -1in
 \advance\evensidemargin
 by -1in
 \marginparsep
 0.4cm
 \advance\topmargin by
 -0.0in
 \makeindex

 \newcommand\la{\langle}
 \newcommand\ra{\rangle}
 \newcommand\beq{\begin{equation}}
 
 \newcommand\eeq{\end{equation}}                                               
 \newcommand\beqn{\begin{eqnarray}}
 \newcommand\eeqn{\end{eqnarray}}
 \newcommand\GeV{{\rm GeV}}

\def\BA{\begin{eqnarray}}
\def\BE{\begin{equation}}
\def\BF{\begin{figure}[htb]}
\def\BT{\begin{table}[htb]}
\def\EA{\end{eqnarray}}
\def\EE{\end{equation}}
\def\EF{\end{figure}}
\def\ET{\end{table}}

\def\la{\langle}
\def\ra{\rangle}
\def\mb{\,\mbox{mb}}
\def\fm{\,\mbox{fm}}
\def\GeV{\,\mbox{GeV}}

\def\lsim{\mathrel{\rlap{\lower4pt\hbox{\hskip1pt$\sim$}}
    \raise1pt\hbox{$<$}}}         %less than or approx. symbol
\def\gsim{\mathrel{\rlap{\lower4pt\hbox{\hskip1pt$\sim$}}
    \raise1pt\hbox{$>$}}}         %greater than or approx. symbol

\begin{document} 
\vspace*{3cm}

\date{today}

\begin{center}
{\LARGE \bf
%%%%%%%%%%%%%%%%%%%%%%%%%%%%%%%%%%%%%%%%%
Nuclear Shadowing in DIS: \\ 
%%%%%%%%%%%%%%%%%%%%%%%%%%%%%%%%%%%%%%%%%
\vspace*{0.1cm}
%%%%%%%%%%%%%%%%%
Numerical Solution of the Evolution
Equation \\
%%%%%%%%%%%%%%%%%%%%%%%%%%%%%%%%%%%%%%%%%
\vspace*{0.3cm}
%%%%%%%%%%%%%%%%%
for the Green Function}  
%\\
%%%%%%%%%%%%%%%%%   
%\vspace*{0.2cm}
%%%%%%%%%%%%%%%%%%%%%%%%%%%%%%%%
%                        } 
%%%%%%%%%%%%%%%%%%%%%%%%%%%%%%%%
\end{center}

%\maketitle

\begin{center}

\vspace{0.5cm}
 {\large J.~Nemchik}
 \\[1cm]
 {\sl Institute of Experimental Physics SAS, Watsonova 47,
04353 Kosice, Slovakia}

\end{center}

\vspace{1cm}
%----------------------------------------------------
\begin{abstract}    

Within a light-cone 
QCD formalism 
based on the Green function
technique incorporating  
color transparency and
coherence length effects 
we study nuclear shadowing in 
deep-inelastic scattering at 
moderately small Bjorken $x_{Bj}$.
Calculations performed so far were
based only on approximations leading to
an analytical harmonic oscillatory form
of the Green function.
We present for the first time
an exact numerical solution of the evolution
equation for the Green function using 
realistic form of the
dipole cross section and nuclear density
function. 
We compare numerical results for nuclear shadowing
with previous predictions and
discuss differences.

\end{abstract}
%----------------------------------------------------
%\doublespace
\newpage

%\pacs{13.85.Lg, 13.60.Le}

%\maketitle

%%%%%%%%%%%%%%%%%%%%%%%%%%%%%%%%%
\section{Introduction} 
\label{intro}
%%%%%%%%%%%%%%%%%%%%%%%%%%%%%%%%%

Nuclear shadowing in deep-inelastic scattering
(DIS) off nuclei is intensively studied
during the last two decades.
It can be treated differently depending
on the reference frame.
In the rest frame of the nucleus
this phenomenon looks like
nuclear shadowing of the hadronic fluctuations
of the virtual photon 
and is occurred due to their multiple scattering
inside the target 
\cite{bsyp-78,fs-88,bl-90,nz-91,mt-93,npz-95,prw-95,kp-96,kp-97,pw-00}.
In the infinite momentum frame of the nucleus
it can be interpreted, however, 
as a result of parton fusion
\cite{k-73,glr-83,mq-86,q-87}
leading to a reduction of the parton density
at low Bjorken $x_{Bj}$. 
Although these two physical interpretations are
complementary, we will work in the
rest frame of the nucleus,
which is more intuitive and is well suited
also for the study of the coherence effects \cite{krt-98}. 

Important phenomenon which controls the dynamics
of nuclear shadowing in DIS is 
effect of quantum coherence.
It results 
from destructive interference of
the amplitudes for which the interaction takes place on different bound
nucleons. It can be treated also as the lifetime of the $\bar qq$
fluctuation and 
estimated by relying on the uncertainty principle and
Lorentz time dilation as,
%
%%%%%%%%%%%%%%%%%%%%%%%%%%%%%%%%%%%
 \beq
t_c = \frac{2\,\nu}{Q^2 + M_{\bar qq}^2}\ ,
%----------
\label{10}
%----------
 \eeq 
%%%%%%%%%%%%%%%%%%%%%%%%%%%%%%%%%%%
%
where $\nu$ is the photon energy, $Q^2$ is photon virtuality
and $M_{\bar qq}$ is the effective mass of the $\bar qq$ pair.
It is usually called coherence time, but we also will use the term
coherence length (CL), since light-cone kinematics is assumed, $l_c=t_c$.
CL is related to the longitudinal momentum transfer $q_c=1/l_c$.
The effect of CL is naturally incorporated in the Green function
formalism already applied in DIS, Drell-Yan pair production 
\cite{krt-98,krt-00} and vector meson production \cite{knst-01,n-02}
(see also the next Section).

The nuclear shadowing in DIS was studied in \cite{krt-98,krt-00}
using correct quantum mechanical treatment based on the Green
function formalism. 
The Green function controls then not only the
relative transverse motion of the $\bar qq$ pair but
also an importance of the higher order multiple scatterings
in the nucleus. 
The solution of the evolution equation for
the Green function was performed so far analytically.
This analytical solution requires, however, to implement
several approximations into a rigorous quantum-mechanical
approach
like a constant nuclear density function (see Eq.~(\ref{190})) and
a specific quadratic form of the dipole cross section (see Eq.~(\ref{180})).
Consequently, 
obtained in a such way the harmonic
oscillator Green function (see Eq.~(\ref{200}))
was used for calculation of nuclear shadowing.
However, the following question naturally arises;
how accurate is the evaluation of the nuclear shadowing
in DIS using this Green function ?
In order to clarify this one should solve
the evolution equation for the Green function numerically.
It does not bring any additional assumptions and
does not force us to use supplementary approximations,
which cause the theoretical uncertainties.
Therefore the main goal of this paper is to 
present for the first time the predictions of
nuclear shadowing in DIS 
at moderately small $x_{Bj}$ based on exact numerical
solution of the evolution equation for the Green function.
In addition, applying an algorithm described in the Appendix~A  
we present also calculations of nuclear shadowing
within the harmonic oscillator Green function approach 
using quadratic form of the dipole cross section (Eq.~(\ref{180}))
and a constant nuclear density function (Eq.~(\ref{190})).
We check whether they correspond to the
results already presented in \cite{krt-98}.
Finally we analyze and discuss the differences between the exact 
and approximate predictions for nuclear shadowing.
Advantages of an exact numerical solution of the 
two-dimensional Schr\" odinger equation 
for the Green function (see Eq.~(\ref{150})) presented in this 
paper provide a better baseline for the future study of the
QCD dynamics not only in DIS off nuclei but also
in further processes occurred in lepton (proton)-nucleus
collisions.

Calculations of nuclear shadowing 
presented in the paper
\cite{krt-98} were performed assuming only
$\bar qq$ fluctuations of the photon and neglecting
higher Fock components containing gluons
and sea quarks.
Performing realistic calculations, 
we include the effects of higher Fock states as the
energy dependence of the dipole cross section, 
$\sigma_{\bar qq}(\vec{r},s)$\footnote{Here $\vec r$
represents the transverse separation of the $\bar qq$ photon
fluctuation
and $s$ is the center of mass energy squared (see the next Section).}.
We use two realistic 
parametrizations of $\sigma_{\bar qq}(\vec{r},s)$ (see the next Section
and Eqs.~(\ref{gbw-1}) and (\ref{kst-1})), where the energy
dependence is naturally included. 
However, we will neglect higher Fock states leading to
gluon shadowing (GS) \cite{kst2} assuming only low and medium
values of the photon energy $\nu$ as was done
also in \cite {krt-98}.

The paper is organized as follows.
In the next Section we present the light-cone dipole 
phenomenology for nuclear shadowing in DIS together
with the Green function formalism.
The Section~\ref{green} supplemented by Appendix~A
is devoted to description of an algorithm
for numerical solution of the evolution equation for the Green
function. 
Numerical results based on realistic
calculations and a comparison with
predictions within harmonic oscillator Green function
approach are presented in Section~\ref{results}.
Finally, in Section~\ref{conclusions} we summarize our main results
and discuss differences between realistic and 
approximate \cite{krt-98,krt-00} calculations of nuclear shadowing in DIS.

%
%%%%%%%%%%%%%%%%%%%%%%%%%%%%%%%%%%%%%%%%%%%%%%%%%%%%%%%%%%%%%%%%%%%%%%%%%%%
\section{Light-cone dipole phenomenology for nuclear
shadowing}
\label{lc}
%%%%%%%%%%%%%%%%%%%%%%%%%%%%%%%%%%%%%%%%%%%%%%%%%%%%%%%%%%%%%%%%%%%%%%%%%%%
%

The main goal of the light-cone (LC) dipole approach to nuclear shadowing 
is a possibility to include the nuclear form factor in all
multiple scattering terms. Derivation of the formula for
nuclear shadowing can be found in \cite{rtv-99}.
The study of the difference between the correct quantum-mechanical
treatment of nuclear shadowing and known approximations
is given in \cite{krt-98} assuming only
$\bar qq$ Fock components of the photon and neglecting
higher Fock components containing gluons and sea quarks. 
The nuclear antishadowing effect was omitted as well
because was assumed to be beyond the shadowing dynamics.
The total photoabsorption cross section on a nucleus can be formally
represented in the form
%
%%%%%%%%%%%%%%%%%%%%%%%%%%%%%%%%%%%%%%%%%%%%%%%%%%%%%%%%%%%
 \BE
\sigma^{\gamma^*A}(x_{Bj},Q^2) =
A~\sigma^{\gamma^*N}(x_{Bj},Q^2) -
\Delta\sigma(x_{Bj},Q^2)\, .
%----------
\label{110}
%----------
 \EE
%%%%%%%%%%%%%%%%%%%%%%%%%%%%%%%%%%%%%%%%%%%%%%%%%%%%%%%%%%%
%
Here the Bjorken variable $x_{Bj}$ is given by
%
%%%%%%%%%%%%%%%%%%%%%%%%%%%%%%%%%%%%%%%%
\beq
x_{Bj} = \frac{Q^2}{2\,m_N\,\nu} \approx 
\frac{Q^2}{Q^2 + s}\, ,
%----------
\label{115}
%----------
\eeq
%%%%%%%%%%%%%%%%%%%%%%%%%%%%%%%%%%%%%%%%
%
where $s$ is the $\gamma^*$-nucleon center
of mass (c.m.) energy squared and 
$m_N$ is mass of the nucleon. 
$\sigma^{\gamma^*N}(x_{Bj},Q^2)$ in (\ref{110}) is 
total photoabsorption
cross section on a nucleon 
%
%%%%%%%%%%%%%%%%%%%%%%%%%%%%%%%%%%%%%%%%%%%%%%%%%%%%%%%%%%%
 \BE
\sigma^{\gamma^*N}(x_{Bj},Q^2) =
\int d^2 r \int_{0}^{1} d\alpha\,\Bigl
| \Psi_{\bar qq}(\vec{r},\alpha,Q^2)\,\Bigr |^2
~\sigma_{\bar qq}(\vec{r},s)\, .
%----------
\label{120}
%----------
 \EE
%%%%%%%%%%%%%%%%%%%%%%%%%%%%%%%%%%%%%%%%%%%%%%%%%%%%%%%%%%%
%
Here $\sigma_{\bar qq}({\vec{r}},s)$ is 
the dipole cross section
which depends on the $\bar qq$
transverse separation $\vec{r}$ and the c.m. energy squared $s$ 
and 
$\Psi_{\bar qq}({\vec{r}},\alpha,Q^2)$ 
is the LC wave function of the $\bar qq$ Fock component of 
the photon which depends also on the
photon virtuality $Q^2$ and the relative share $\alpha$ of the photon
momentum carried by the quark. 
Note that Bjorken $x_{Bj}$ is related with
c.m. energy squared $s$ via Eq.~(\ref{115}).
Consequently, hereafter we will write the energy dependence
of variables in subsequent formulas also via $x_{Bj}$- dependence whenever
convenient.

The first ingredient of the photoabsorption cross section
on a nucleon (\ref{120}) is
the dipole cross section $\sigma_{\bar qq}(\vec r,s)$ 
representing interaction of a
$\bar qq$ dipole of transverse separation $\vec r$ with a nucleon
\cite{zkl}.
It is a flavor independent universal function of
$\vec{r}$ and energy and allows to describe 
various high energy processes
in an uniform way.
It is known to vanish quadratically
$\sigma_{\bar qq}(r,s)\propto r^2$ as $r\rightarrow 0$ due to color
screening (property of color transparency \cite{zkl,bbgg,bm-88}) and
cannot be predicted
reliably because of poorly known higher order 
perturbative QCD (pQCD) corrections and
nonperturbative effects. 
There are two popular parameterizations of $\sigma_{\bar qq}(\vec 
r,s)$, GBW presented in \cite{gbw} and KST suggested in \cite{kst2}.
Detailed discussion and comparison of these two parametrizations  
can be found for example in \cite{r-00,knst-01}.
Therefore, for completeness, we present here only the main
features of both parametrizations 
because they are used in the realistic
calculations of nuclear shadowing in DIS with the results
shown in Section~\ref{results}.

The GBW model \cite{gbw}
for the dipole cross section
provides a very simple parametrization which saturates at large
$\bar qq$ separations,
%
%%%%%%%%%%%%%%%%%%%%%%%%%%%%%%%%
\beq
\sigma_{\bar qq}(r,x_{Bj}) = \sigma_0\,\left [1 - 
exp\left ( - \frac{r^2}{R_0^2(x_{Bj})}\right )\right ]\, ,
%-----------
\label{gbw-1}
%-----------
\eeq
%%%%%%%%%%%%%%%%%%%%%%%%%%%%%%%%
%
where $R_0(x_{Bj}) = 0.395\,(x_{Bj}/x_0)^{\lambda/2}\fm$ and 
$\sigma_0 = 23.03\mb$; $\lambda = 0.288$; $x_0 = 0.0003$.
This dipole cross section vanishes $\propto r^2$ at small
dipole sizes as implied by color transparency (CT).
It describes well the data for DIS at small 
$x_{Bj}$ and medium and large $Q^2$. However, it cannot be
correct at small $Q^2$ since predicts energy-independent
hadronic cross sections. Besides, $x_{Bj}$ is not any more
a proper variable at small $Q^2$ and should be replaced
by energy. 
This problem is removed by the KST parametrization
\cite{kst2} which keeps the form (\ref{gbw-1}) but
contains an explicit dependence on energy,
%
%%%%%%%%%%%%%%%%%%%%%%%%%%%%%%%%
\beq
\sigma_{\bar qq}(r,s) = \sigma_0(s)\,\left [1 - 
exp\left ( - \frac{r^2}{R_0^2(s)}\right )\right ]\, .
%-----------
\label{kst-1}
%-----------
\eeq
%%%%%%%%%%%%%%%%%%%%%%%%%%%%%%%%
%
An explicit energy dependence in the parameter $\sigma_0(s)$
is introduced in a such way that guarantees the reproduction
of the correct hadronic cross sections,
%
%%%%%%%%%%%%%%%%%%%%%%%%%%%%%%%%%
\beq
\sigma_0(s) = \sigma_{tot}^{\pi\,p}(s)\,\left 
(1 + \frac{3\,R_0^2(s)}{8\,\la r_{ch}^2\ra_{\pi}}\right )\, ,
%------------
\label{kst-2}
%------------
\eeq
%%%%%%%%%%%%%%%%%%%%%%%%%%%%%%%%%
%
where $\sigma_{tot}^{\pi\,p}(s) = 23.6\,(s/s_0)^{0.079} + 
1.432\,(s/s_0)^{-0.45}\mb$ is the Pomeron and Reggeon parts  of the
$\pi p$ total cross section \cite{rpp-96},
and $R_0(s) = 0.88\,(s/s_0)^{-\lambda/2}\fm$ with $\lambda = 0.28$
and $s_0 = 1000\GeV^2$ is the energy-dependent radius.
In Eq.~(\ref{kst-2}) $\la r_{ch}^2\ra_{\pi} = 0.44\fm^2$
is the mean pion charge radius squared.
The main advantage of
the KST parametrization (\ref{kst-2}) is that it describes well the 
transition down to limit of real photoproduction, $Q^2=0$. 
However, the improvement compared to GBW model \cite{gbw}
at large separations (small values of $Q^2$) 
leads to a worse description of the
short-distance part of the dipole cross section which is
responsible for the behavior of the proton structure function
at large $Q^2$. To satisfy Bjorken scaling, the dipole cross
section at small dipole sizes $r$ must be a function
of the product $s\,r$ which is not the case for the KST
parametrization (\ref{kst-1}). The form of Eq.~(\ref{kst-1})
successfully describes the data for DIS at small $x_{Bj}$
only up to $Q^2\approx 10\GeV^2$ and does a poor
job at larger values of $Q^2$.

Summarizing, the GBW model is suited better at medium and large 
$Q^2\gsim 5\div 10\GeV^2$\footnote{i.e.
at medium small and small values of dipole size, $r\propto 
\sqrt{1/Q^2}\lsim 0.06\div 0.09\fm$.}
and at medium small and small $x_{Bj}\lsim 0.01$
whereas the KST model prefers low and medium values of 
$Q^2\lsim 5\div 10\GeV^2$.
Therefore, the difference of the realistic calculations 
for nuclear shadowing in DIS using these two models
for the dipole cross section in the common kinematic region
of their applicability can be treated as a measure
of theoretical uncertainty. 

The second ingredient of $\sigma^{\gamma^*N}(x_{Bj},Q^2)$ 
in (\ref{120}) is
the perturbative distribution amplitude (``wave function'') of the $\bar
qq$ Fock component of the photon\footnote{We
neglect the nonperturbative effects responsible
for the interaction between $\bar q$ and $q$ assuming
sufficiently large values of $Q^2$ in DIS (see below).}
and 
has the following form 
for transversely (T) and longitudinally (L) polarized photons 
\cite{lc,bks-71,nz-91},
%
%%%%%%%%%%%%%%%%%%%%%%%%%%%%%%%%%%%%%%%%%%%%%%
 \BE
\Psi_{\bar qq}^{T,L}({\vec{r}},\alpha,Q^2) =
\frac{\sqrt{N_{C}\,\alpha_{em}}}{2\,\pi}\,\,
Z_{q}\,\bar{\chi}\,\hat{O}^{T,L}\,\chi\, 
K_{0}(\epsilon\,r)
%----------
\label{122}
%----------
 \EE
%%%%%%%%%%%%%%%%%%%%%%%%%%%%%%%%%%%%%%%%%%%%%%
%
where $\chi$ and $\bar{\chi}$ are the spinors of the quark and
antiquark, respectively; $Z_{q}$ is the quark charge,
$N_{C} = 3$ is the
number of colors. $K_{0}(\epsilon r)$ is a modified Bessel
function with 
%
%%%%%%%%%%%%%%%%%%%%%%%%%%%%%%%%%%%%%%%%%%%%
 \BE
\epsilon^{2} =
\alpha\,(1-\alpha)\,Q^{2} + m_{q}^{2}\ ,
%----------
\label{123}
%----------
 \EE
%%%%%%%%%%%%%%%%%%%%%%%%%%%%%%%%%%%%%%%%%%%%
%
where $m_{q}$ is the quark mass.
The operators
$\widehat{O}^{T,L}$ read,
%
%%%%%%%%%%%%%%%%%%%%%%%%%%%%%%%%%%%%%%%%%%%%%%%%%%%%%%%%%%%%
 \BE 
\widehat{O}^{T} = m_{q}\,\,\vec{\sigma}\cdot\vec{e} +
i\,(1-2\alpha)\,(\vec{\sigma}\cdot\vec{n})\,
(\vec{e}\cdot\vec{\nabla}_r) + (\vec{\sigma}\times
\vec{e})\cdot\vec{\nabla}_r\ ,
%-----------
 \label{124}
%-----------
 \EE
%%%%%%%%%%%%%%%%%%%%%%%%%%%%%%%%%%%%%%%%%%%%%%%%%%%%%%%%%%%%
%
%
%%%%%%%%%%%%%%%%%%%%%%%%%%%%%%%%%%%%%%%%%%%%%%%%%%%%%%%%%%%%
 \BE
\widehat{O}^{L} =
2\,Q\,\alpha (1 - \alpha)\,(\vec{\sigma}\cdot\vec{n})\ .
%----------
\label{125}
%----------
 \EE
%%%%%%%%%%%%%%%%%%%%%%%%%%%%%%%%%%%%%%%%%%%%%%%%%%%%%%%%%%%%
%
 Here $\vec\nabla_r$ acts on transverse coordinate $\vec r$;
$\vec{e}$ is the polarization vector of the photon, $\vec{n}$ is a unit
vector parallel to the photon momentum and $\vec{\sigma}$ is the three
vector of the Pauli spin-matrices.

Matrix element (\ref{120}) contains the LC wave function squared,
which has the following form for T and L polarization,
%
%%%%%%%%%%%%%%%%%%%%%%%%%%%%%%%%%%%%%%%%%%%%%%%%%%%%%%%%%%%
 \BE
\Bigl |\Psi^{T}_{\bar qq}(\vec r,\alpha,Q^2)\,\Bigr |^2 =
\frac{2\,N_C\,\alpha_{em}}{(2\pi)^2}\,
\sum_{f=1}^{N_f}\,Z_f^2
\left[m_f^2\,K_0(\epsilon,r)^2
+ [\alpha^2+(1-\alpha)^2]\,\epsilon^2\,K_1(\epsilon\,r)^2\right]\ ,
%----------
\label{127a}
%----------
 \EE
%%%%%%%%%%%%%%%%%%%%%%%%%%%%%%%%%%%%%%%%%%%%%%%%%%%%%%%%%%%
%
and
%
%%%%%%%%%%%%%%%%%%%%%%%%%%%%%%%%%%%%%%%%%%%%%%%%%%%%%%%%%%%
 \BE
\Bigl |\Psi^{L}_{\bar qq}(\vec r,\alpha,Q^2)\,\Bigr |^2 =
\frac{8\,N_C\,\alpha_{em}}{(2\pi)^2}\,
\sum_{f=1}^{N_f}\,Z_f^2
\,Q^2\,\alpha^2(1-\alpha)^2\,
K_0(\epsilon\,r)^2\ ,
%----------
\label{127b}
%----------
 \EE
%%%%%%%%%%%%%%%%%%%%%%%%%%%%%%%%%%%%%%%%%%%%%%%%%%%%%%%%%%%
%   
where $K_1$ is the modified Bessel function,
%
%%%%%%%%%%%%%%%%%%%%%%%%%%%%%%%%%%%%%%%%%%%%%%%%%%%%%%%%%%%
 \BE
K_1(z) = - \frac{d}{dz}\,K_0(z)\, .
%----------
\label{128}
%----------
 \EE
%%%%%%%%%%%%%%%%%%%%%%%%%%%%%%%%%%%%%%%%%%%%%%%%%%%%%%%%%%%
%   

Note that in the LC formalism the photon wave function contains
also higher Fock states $|\bar qq\ra$, $|\bar qqG\ra$, 
$|\bar qq2G\ra$, etc.
The effects of higher Fock states are implicitly incorporated
into the energy 
(Bjorken $x_{Bj}$- ) dependence of the dipole cross section
$\sigma_{\bar qq}(\vec{r},s)$ as is given in Eq.~(\ref{120}).
Note, that the energy dependence of the dipole cross section
is naturally included in realistic parametrizations Eqs.~(\ref{gbw-1})
and (\ref{kst-1}). 

In Eq.~(\ref{110})
the second term, $\Delta\sigma$,
represents the shadowing correction
and has the following form
%
%%%%%%%%%%%%%%%%%%%%%%%%%%%%%%%%%%%%%%%%%%%%%%%%%%%%%%%%%%%
 \BE
\Delta\sigma(x_{Bj},Q^2) = \frac{1}{2}~{Re}~\int d^2 b \int_{-\infty}^{
\infty} dz_1 ~\rho_{A}(b,z_1) \int_{z_1}^{\infty} dz_2~
\rho_A(b,z_2) 
\int_{0}^{1} d\alpha ~A(z_1,z_2,\alpha)\, ,
%----------
\label{130}
%----------
 \EE
%%%%%%%%%%%%%%%%%%%%%%%%%%%%%%%%%%%%%%%%%%%%%%%%%%%%%%%%%%%
%
with 
%
%%%%%%%%%%%%%%%%%%%%%%%%%%%%%%%%%%%%%%%%%%%%%%%%%%%%%%%%%%%
 \BE
A(z_1,z_2,\alpha) 
= \int d^2 r_2 ~\Psi^{*}_{\bar qq}(\vec{r_2},\alpha,Q^2)
~\sigma_{\bar qq}(r_2,s) \int d^2 r_1
~G_{\bar qq}(\vec{r_2},z_2;\vec{r_1},z_1)
~\sigma_{\bar qq}(r_1,s)
~\Psi_{\bar qq}(\vec{r_1},\alpha,Q^2)\, .
%----------
\label{140}
%----------
 \EE
%%%%%%%%%%%%%%%%%%%%%%%%%%%%%%%%%%%%%%%%%%%%%%%%%%%%%%%%%%%
%
In Eq.~(\ref{130}) 
$\rho_{A}({b},z)$ represents the nuclear density function defined
at the point with longitudinal coordinate $z$ and impact
parameter $\vec{b}$. 
%
%****************************************************************
%************************ FIG.1 *********************************
%****************************************************************
 \begin{figure}[tbh]
\includegraphics{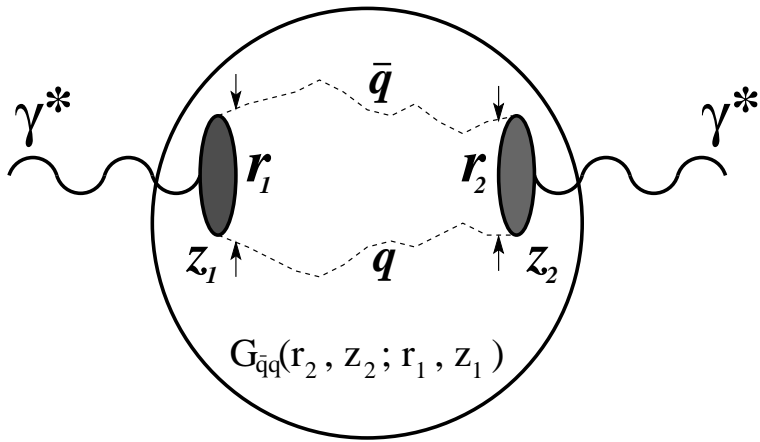}
\begin{center}
\vspace{6.7cm}
\parbox{13cm}
{\caption[Delta]
 {A cartoon for the shadowing term $\Delta\sigma$ in (\ref{110}).
Propagation of the $\bar qq$ pair through the nucleus is
described by the Green function 
$G_{\bar qq}(\vec{r_2},z_2;\vec{r_1},z_1)$ which results
from the summation over different paths of the $\bar qq$ pair.}
%%%%%%%%%%%%%%%%%%%%%%%%%
 \label{shad}}
%%%%%%%%%%%%%%%%%%%%%%%%%
\end{center}
 \end{figure}
%****************************************************************
%

The shadowing term in (\ref{110}) is illustrated in Fig.~\ref{shad}.
At the point $z_1$ the initial photon diffractively produces
the $\bar qq$ pair ($\gamma^*N\to \bar qqN$) with 
transverse separation $\vec{r_1}$. The $\bar qq$ pair then propagates
through the nucleus along arbitrary curved trajectories,
which are summed over, and arrives at the point $z_2$ with
a transverse separation $\vec{r_2}$.
The initial and final separations are controlled by the  
LC wave function of the $\bar qq$ Fock component of
the photon $\Psi_{\bar qq}(\vec{r},\alpha,Q^2)$.
During propagation through the nucleus the $\bar qq$
pair interacts with bound nucleons via dipole cross section
$\sigma_{\bar qq}(r,s)$ which depends on the local
transverse separation $\vec{r}$. The Green function
$G_{\bar qq}(\vec{r_2},z_2;\vec{r_1},z_1)$ describes 
the propagation of the $\bar qq$ pair from $z_1$ to $z_2$.

The Green function $G_{\bar qq}(\vec{r_2},z_2;\vec{r_1},z_1)$ 
satisfies the time-dependent
two-dimensional Schr\"odinger equation, 
%
%%%%%%%%%%%%%%%%%%%%%%%%%%%%%%%%%%%%%%%%%%%%%%%%%%%%%%%%%%%%%%%%%%%%%%
 \BE
i\frac{d}{dz_2}\,G_{\bar qq}(\vec{r_2},z_2;\vec{r_1},z_1)=
\left[\frac{\epsilon^{2} - \Delta_{r_{2}}}{2\,\nu\,\alpha\,(1-\alpha)}
+V_{\bar qq}(z_2,\vec r_2,\alpha)\right]
G_{\bar qq}(\vec{r_2},z_2;\vec{r_1},z_1)\ ,
%----------
\label{150}
%----------  
 \EE
%%%%%%%%%%%%%%%%%%%%%%%%%%%%%%%%%%%%%%%%%%%%%%%%%%%%%%%%%%%%%%%%%%%%%%
%
with the boundary condition
%
%%%%%%%%%%%%%%%%%%%%%%%%%%%%%%%%
\beq
G_{\bar qq}(\vec{r_2},z_2;\vec{r_1},z_1)|_{z_2=z_1}=
\delta^2(\vec r_1-\vec r_2)\, .
%----------
\label{155}
%----------
\eeq
%%%%%%%%%%%%%%%%%%%%%%%%%%%%%%%%
%
In Eq.~(\ref{150})  
the Laplacian $\Delta_{r}$ acts on
the coordinate $r$ and $\epsilon$ is given by (\ref{123}).

The Green function 
$G_{\bar qq}(\vec{r_2},z_2;\vec{r_1},z_1)$
includes the phase shift
between initial and final photons which is due to transverse
and also longitudinal motion of the quarks.
One can see the presence of the CL (\ref{10}) in the kinetic term
of the evolution Schr\"odinger equation (\ref{150}),
where the role of time is played by the
longitudinal coordinate $z_2$.
A part $\Delta_r/[2\nu\alpha(1-\alpha)]$
of this kinetic term 
takes care of the varying effective
mass of the $\bar qq$ pair,
$M_{\bar qq}^2 = (m_q^2 + k_T^2)/\alpha(1-\alpha)$,
and provides a proper phase shift. 
This is what the overall kinetic term consists when
the transverse momentum squared of the quark is replaced
by $k_T^2\rightarrow \Delta_r$.
This dynamically varying effective mass controls CL defined by the
Green function.
The static part $Q^2 + m_q^2/\alpha(1-\alpha)$
of the CL is connected with
the longitudinal motion and is included
in the Green function as well via the last phase shift
factor (see Eq.~(\ref{200}) below). 
Consequently, the longitudinal momentum transfer is known
and all the multiple interactions are included.

The imaginary part of the LC potential $V_{\bar
qq}(z_2,\vec r_2,\alpha)$ in Eq.~(\ref{150}) is responsible for
attenuation of the $\bar qq$ photon fluctuation in the medium, 
while the real part
represents the interaction between the $q$ and $\bar{q}$.
Because we are going to calculate the nuclear shadowing
in DIS at medium and large $Q^2$ one can safely neglect
this $\bar q$-$q$ interaction as was done also in \cite{krt-98}.

In the LC Green function approach \cite{krt-98,krt-00,knst-01,n-02}
the physical photon 
$|\gamma^*\ra$ is decomposed 
into different Fock states, namely, the bare photon
$|\gamma^*\ra_0$, $|\bar qq\ra$, $|\bar qqG\ra$, etc. 
As we mentioned above the higher Fock states
containing gluons describe the energy dependence of the
photoabsorption cross section on a nucleon. 
Besides, those Fock components lead also to GS  
as far as nuclear effects are concerned.
However, these fluctuations are heavier and have a
shorter coherence time (lifetime) than the lowest $|\bar qq\ra$
state. Therefore, at medium energies only $|\bar qq\ra$ fluctuations of 
the photon matter. Consequently, GS related to the higher Fock
states will be dominated at high energies.
Because we will calculate the nuclear shadowing at moderately small
$x_{Bj}$ (medium values of $\nu$) we can neglect the
GS for our purposes.
This is supported also by the main goal of
this paper which is based on comparison
of the realistic calculations for nuclear shadowing with the results
obtained within the harmonic oscillatory Green function approach and 
presented in the paper \cite{krt-98} where GS was neglected as well. 

One can describe a propagation of a noninteracting $\bar qq$ pair in a 
nuclear medium 
by the Green function satisfying the evolution Eq.~(\ref{150}).
The LC potential $V_{\bar qq}(z_2,\vec r,\alpha)$  
in this case acquires only an imaginary part which
represents absorption in the medium,
%
%%%%%%%%%%%%%%%%%%%%%%%%%%%%%%%%%%%%%%%%%%%%%%%%%%%%%%%%%%%%
 \BE
Im V_{\bar qq}(z_2,\vec r,\alpha) = - 
\frac{\sigma_{\bar qq}(\vec r,s)}{2}\,\rho_{A}({b},z_2)\, .
%----------
\label{170}
%----------
 \EE
%%%%%%%%%%%%%%%%%%%%%%%%%%%%%%%%%%%%%%%%%%%%%%%%%%%%%%%%%%%%
%

The analytical solution of Eq.~(\ref{150}) is known only for the 
harmonic oscillator potential $V(r)\propto r^2$. 
Consequently, one should use the 
dipole approximation
%
%%%%%%%%%%%%%%%%%%%%%%%%%%%%%%%%%%%%%%%%%%
 \beq
\sigma_{\bar qq}(r,s) = C(s)\,r^2\ ,
%----------
\label{180}
%----------
 \eeq
%%%%%%%%%%%%%%%%%%%%%%%%%%%%%%%%%%%%%%%%%%
%
and uniform nuclear density 
%
%%%%%%%%%%%%%%%%%%%%%%%%%%%%%%%%%%%%%%%%%%
 \beq
\rho_A(b,s) = \rho_0~\Theta(R_A^2-b^2-z^2)\, ,
%----------
\label{190}
%----------
 \eeq
%%%%%%%%%%%%%%%%%%%%%%%%%%%%%%%%%%%%%%%%%%
%
in order to to obtain the Green function in an analytical form.
In Eq.~(\ref{190}) $R_A$ is the nuclear radius.
The solution in this case is the
harmonic oscillator Green function \cite{fg},
%
%%%%%%%%%%%%%%%%%%%%%%%%%%%%%%%%%%%%%%%%%%%%%%%%%%%%%%%%%%%%
 \BA 
G_{\bar qq}(\vec{r_2},z_2;\vec{r_1},z_1) &=&
\frac{b(\alpha)}{2\,\pi\,i\,
{\rm sin}(\omega\,\Delta z)}\, {\rm exp}
\left\{\frac{i\,b(\alpha)}{{\rm sin}(\omega\,\Delta z)}\,
\Bigl[(r_1^2+r_2^2)\,{\rm cos}(\omega \,\Delta z) -
2\,\vec r_1\cdot\vec r_2\Bigr]\right\}
\nonumber\\ \times \,
& &{\rm exp}\left[- 
\frac{i\,\epsilon^{2}\,\Delta z}
{2\,\nu\,\alpha\,(1-\alpha)}\right] \ , 
%----------
\label{200} 
%----------
 \EA
%%%%%%%%%%%%%%%%%%%%%%%%%%%%%%%%%%%%%%%%%%%%%%%%%%%%%%%%%%%%
%
%
where $\Delta z=z_2-z_1$ and
%
%%%%%%%%%%%%%%%%%%%%%%%%%%%%%%%%%%%%%%%%%%%%%%%%%%%%%%%%%%
 \BE 
\omega = \frac{b(\alpha)}{\nu\;\alpha(1-\alpha)}\ ,
%----------
\label{210}
%---------- 
 \EE
%%%%%%%%%%%%%%%%%%%%%%%%%%%%%%%%%%%%%%%%%%%%%%%%%%%%%%%%%%
%
where
%
%%%%%%%%%%%%%%%%%%%%%%%%%%%%%%%%%%%%%%%%%%%%%%%%%%%%%
 \beq
b^2(\alpha) = - i\,\rho_{A}({b},z)\,
\nu\,\alpha\,(1-\alpha)\,C(s)\ .
%----------
\label{220}
%----------
 \eeq
%%%%%%%%%%%%%%%%%%%%%%%%%%%%%%%%%%%%%%%%%%%%%%%%%%%%%
%

The energy dependent factor $C(s)$ in Eq.~(\ref{180}) 
and the mean nuclear density 
$\rho_0$ in Eq.~(\ref{190}) can be adjusted 
by the procedure described in \cite{krt-00}.
According this procedure the factor $C(s)$ is fixed by demanding that
calculations employing the approximation Eq.~(\ref{180})
reproduce correctly the results based on the realistic
cross section (given by KST parametrization Eq.~(\ref{kst-1}) or 
by GBW model Eq.~(\ref{gbw-1})) in the high-energy limit $l_c\gg R_A$
when the Green function takes the simple form 
(see Eq.~(\ref{250}) below).
Consequently, the factor $C(s)$ is fixed by the relation,
%
%%%%%%%%%%%%%%%%%%%%%%%%%%%%%%%%%%%%%%%%%%%%%%%%%%%
\BA
\frac
{\int d^2\,b\,\int d^2\,r\,\Bigl 
|\Psi_{\bar qq}^{T,L}(\vec{r},\alpha,Q^2)\Bigr |^2\,
\left\{1 - \exp\,\Bigl[ - \frac{1}{2}\,C(s)\,r^2\, 
T_A(b)\Bigr]\,\right\}} 
{\int d^2\,r\,
\Bigl |\Psi_{\bar qq}^{T,L}(\vec{r},\alpha,Q^2)\Bigr |^2\,
C(s)\,r^2}
\nonumber \\
=
\frac{\int d^2\,b\,\int d^2\,r\,\Bigl 
|\Psi_{\bar qq}^{T,L}(\vec{r},\alpha,Q^2)\Bigr |^2\,
\left\{1 - \exp\,\Bigl[ - \frac{1}{2}\,\sigma_{\bar qq}(r,s)\, 
T_A(b)\Bigr]\,\right\}}
{\int d^2\,r\,
\Bigl |\Psi_{\bar qq}^{T,L}(\vec{r},\alpha,Q^2)\Bigr |^2\,
\sigma_{\bar qq}(r,s)}\, ,
%-----------
\label{224}
%-----------
\EA
%%%%%%%%%%%%%%%%%%%%%%%%%%%%%%%%%%%%%%%%%%%%%%%%%%%
%
where
%
%%%%%%%%%%%%%%%%%%%%%%%%%%%
\beq
T_A(b) = \int_{-\infty}^{\infty}\,dz\,\rho_A(b,z)\,
%-----------
\label{226}
%-----------
\eeq
%%%%%%%%%%%%%%%%%%%%%%%%%%%
%
is the nuclear thickness calculated with the realistic Wood-Saxon
form of the nuclear density with parameters taken from \cite{saxon}.
This procedure is performed separately for T and L
polarized photons and for each value of 
$\alpha$. 
The value for the mean nuclear density $\rho_0$ in Eq.~(\ref{190})
is determined in a similar way using relation
%
%%%%%%%%%%%%%%%%%%%%%%%%%%%
\beq
\int\,d^2\,b\,\Biggl [1 - exp\,\Biggl ( - \sigma_0\,\rho_0
\,\sqrt{R_A^2 - b^2}\,\Biggr )\,\Biggr ] =
\int\,d^2\,b\,\Biggl [1 - exp\,\Biggl ( - 
\frac{1}{2}\,\sigma_0\,T_A(b)\,\Biggr )\,\Biggr ]\, .
%-----------
\label{228}
%-----------
\eeq
%%%%%%%%%%%%%%%%%%%%%%%%%%%
%
The value of $\rho_0$ turns out to be practically independent
of the cross section $\sigma_0$ from 1 to 50$\mb$ as was
checked in \cite{krt-00,r-00}.

We would like to emphasize that only in the high energy
limit, $l_c\gg R_A$, it is possible to resum
the whole multiple scattering series in an eikonal-formula.
Correspondingly, the transverse separation $r$ between
$\bar q$ and $q$ does not vary 
during propagation through the nucleus (Lorentz time dilation).
Then the total photoabsorption cross section on a nucleus reads
\cite{zkl}
%
%%%%%%%%%%%%%%%%%%%%%%%%%%%%%%%%%%%%%%%%%%%%%%%%%%%%%%%%%%%%
 \BA
\sigma^{\gamma^*A}(s,Q^2) &=&
2\,\int d^2b\,\int d^2r\,\int_{0}^{1}
d\alpha\,|\Psi_{\bar qq}(\vec{r},\alpha,Q^2)|^2\, 
\left\{1 - \exp\,\Bigl[ - \frac{1}{2}\,\sigma_{\bar qq}(r,s)\, 
T_A(b)\Bigr]\,\right\} \nonumber\\
&\equiv&
2\,\int d^2b\,
\left\{1 - \Biggl\la\,\exp\,\Bigl[ - \frac{1}{2}\,\sigma_{\bar 
qq}(r,s)\, 
T_A(b)\Bigr]\,\Biggr\ra\,\right\}\, .
%----------
\label{230} 
%----------
 \EA
%%%%%%%%%%%%%%%%%%%%%%%%%%%%%%%%%%%%%%%%%%%%%%%%%%%%%%%%%%%%
%
Note that the averaging of the whole exponential in (\ref{230})
makes this expression different from the Glauber eikonal
approximation where $\sigma(r,s)$ is averaged in the exponent,
%
%%%%%%%%%%%%%%%%%%%%%%%%%%%%%%%%%%%%%%%%%%%%%%%%%%%%%%%%%%%%
 \BA
\sigma^{\gamma^*A}_{Glauber}(s,Q^2) =
2\,\int d^2b\,
\left\{1 - \,\exp\,\Bigl[ - \frac{1}{2}\,\sigma^{\gamma^*N}(s,Q^2)\, 
T_A(b)\Bigr]\,\right\}\, .
%----------
\label{240} 
%----------
 \EA
%%%%%%%%%%%%%%%%%%%%%%%%%%%%%%%%%%%%%%%%%%%%%%%%%%%%%%%%%%%%
%
The difference is known as Gribov's inelastic corrections
\cite{gribov}.
In the case of DIS the Glauber approximation does not
make sense (because of a small value of $\sigma^{\gamma^*p}$,
which is at most of the order of $100\,\mu b$ for real photons)
and the whole cross section is due to the inelastic
shadowing.

The eikonal formula (\ref{230}) for the total photoabsorption
cross section on a nucleus can be obtained as a limiting case
of the Green function formalism.
Indeed,
in the high energy limit, $\nu\rightarrow \infty$ the kinetic
term in Eq.~(\ref{150}) can be neglected and the Green function
reads
%
%%%%%%%%%%%%%%%%%%%%%%%%%%%%%%%%%%%%%%%%%%%%%%%%%%%%%
\BA
G_{\bar qq}(\vec{r_2},z_2;\vec{r_1},z_1)|_{\nu\to\infty} = 
\delta(\vec{r_2}-\vec{r_1})\,\exp\Biggl[ - \frac{1}{2}\,
\sigma_{\bar qq}(r_2,s)\,\int_{z_1}^{z_2}\,dz\,\rho_A(b,z)\Biggr]\,.
%----------
\label{250}
%----------
\EA
%%%%%%%%%%%%%%%%%%%%%%%%%%%%%%%%%%%%%%%%%%%%%%%%%%%%%
%
After substitution of this expression into Eqs.~(\ref{110}),
(\ref{130}) and (\ref{140}) one arrives at the result
(\ref{230}). 

For smaller energies when $l_c\sim R_A$, 
one has to take into account the variation of the transverse size
$r$ during propagation of the $\bar qq$ pair through the nucleus.
This transverse size variation is naturally included 
using correct quantum-mechanical
treatment based on the Green function formalism presented above.

The overall total photoabsorption cross section on a nucleus
is given as a sum of T and L polarizations,
$\sigma^{\gamma^*A} = \sigma_T^{\gamma^*A} +
\epsilon'\,\sigma_L^{\gamma^*A}$, assuming that the photon
polarization $\epsilon'=1$.
If one takes into account only $\bar qq$ Fock component
of the photon
the full expression after summation over all flavors, colors,
helicities and spin states has the following form \cite{bgz-98}
%
%%%%%%%%%%%%%%%%%%%%%%%%%%%%%%%%%%%%%%%%%%%%%%%%%%%%%%%
\BA
\sigma^{\gamma^*A}(x_{Bj},Q^2) &=&
A\,\sigma^{\gamma^*N}(x_{Bj},Q^2) - \Delta\,\sigma(x_{Bj},Q^2) 
\nonumber \\
&=& A\,\int\,d^2r\,\int_{0}^{1}\,d\alpha\,\sigma_{\bar qq}(r,s)
\,\Biggl (\Bigl |\Psi^T_{\bar qq}(\vec{r},\alpha,Q^2)\Bigr |^2 +
\Bigl |\Psi^L_{\bar qq}(\vec{r},\alpha,Q^2)\Bigr |^2\Biggr )
\nonumber \\
&-& \frac{3\,\alpha_{em}}{(2\pi)^2}\,\sum_{f=1}^{N_f}\,Z_f^2\,Re\,
\int\,d^2b\,\int_{-\infty}^{\infty}\,dz_1\,\int_{z_1}^{\infty}\,
dz_2\,\int_{0}^{1}\,d\alpha\,\int\,d^2r_1\,\int\,d^2r_2
\nonumber \\
&&\times\,\rho_A(b,z_1)\,\rho_A(b,z_2)\,\sigma_{\bar qq}(r_2,s)\,
\sigma_{\bar qq}(r_1,s)
\nonumber \\
&&\times\,\Biggl\{\Bigl[\,\alpha^2 + (1 - \alpha)^2\,\Bigr]
\,\epsilon^2
\,\frac{\vec{r_1}\,\cdot\,\vec{r_2}}{r_1\,r_2}\,
K_1(\epsilon\,r_1)\,K_1(\epsilon\,r_2)
%----------
\label{260}
%----------
\\
&&\,\,\,\,\,\,\,\,\,\,\,\,\,\,\,
 + \,\Bigl[\,m_f^2 + 4\,Q^2\,\alpha^2\,(1 - \alpha)^2\,\Bigr]\,
K_0(\epsilon\,r_1)\,K_0(\epsilon\,r_2)\Biggr\}\,
G_{\bar qq}(\vec{r_2},z_2;\vec{r_1},z_1) \, .
\nonumber
\EA
%%%%%%%%%%%%%%%%%%%%%%%%%%%%%%%%%%%%%%%%%%%%%%%%%%%%%%%
%
Here $\Bigl |\,\Psi^{T,L}_{\bar qq}(\vec{r},\alpha,Q^2)\,\Bigr |^2$
are the absolute squares of the LC wave functions for
the $\bar qq$ fluctuation of T and L
polarized photons summed over all flavors
with the form given by Eqs.~(\ref{127a})
and (\ref{127b}), respectively.

In the high energy limit after substitution of expression
(\ref{250}) for the Green function into Eq.~(\ref{260})
one arrives at the following results, which corresponds
to Eq.~(\ref{230}) after inclusion of a sum of T and L
polarizations :
%
%%%%%%%%%%%%%%%%%%%%%%%%%%%%%%%%%%%%%%%%%%%%%%%%%%%%%%%
\BA
\sigma^{\gamma^*A}(x_{Bj},Q^2) &=&
2\,\int\,d^2b\,\int\,d^2r\,\int_0^1\,d\alpha
\left\{1 - \exp\,\Bigl[ - \frac{1}{2}\,\sigma_{\bar qq}(r,s)\, 
T_A(b)\Bigr]\,\right\} \nonumber\\
&&\times\,\frac{2\,N_C\,\alpha_{em}}{(2\pi)^2}\,\sum_{f=1}^{N_f}\,Z_f^2\,
\Biggl\{\Bigl[\,\alpha^2 + (1 - \alpha)^2\,\Bigr]
\,\epsilon^2\,K_1^2(\epsilon\,r)\,
%----------
\label{265}
%----------
\\
&&\,\qquad\qquad\qquad\qquad
 + \,\Bigl[\,m_f^2 + 4\,Q^2\,\alpha^2\,(1 - \alpha)^2\,\Bigr]\,
K_0^2(\epsilon\,r)\,\Biggr\}\, .
\nonumber
\EA
%%%%%%%%%%%%%%%%%%%%%%%%%%%%%%%%%%%%%%%%%%%%%%%%%%%%%%%
%

At photon polarization parameter $\epsilon'=1$
the structure function ratio 
$F_2^A(x_{Bj},Q^2)/F_2^N(x_{Bj},Q^2)$ can be expressed via
ratio of the total photoabsorption cross sections
%
%%%%%%%%%%%%%%%%%%%%%%%%%%%%%%
\BA
\frac{F_2^A(x_{Bj},Q^2)}{F_2^N(x_{Bj},Q^2)} = 
\frac{\sigma_T^{\gamma^*A}(x_{Bj},Q^2)
+ \sigma_L^{\gamma^*A}(x_{Bj},Q^2)}
{\sigma_T^{\gamma^*N}(x_{Bj},Q^2)
+ \sigma_L^{\gamma^*N}(x_{Bj},Q^2)}\, ,
%----------
\label{270}
%----------
\EA
%%%%%%%%%%%%%%%%%%%%%%%%%%%%%%
%
where the numerator on right-hand side (r.h.s.) is given by 
Eq.~(\ref{260}), whereas denominator can be expressed
as the first term of Eq.~(\ref{260})
divided by the mass number $A$.

Finally we would like to emphasize that $\bar qq$ Fock component 
of the photon represents the higher twist shadowing correction 
\cite{krt-00}.
This correction vanishes at large quark masses as $1/m_f^2$.
Not so for the higher Fock states containing gluons and leading to GS.
GS represents the leading twist shadowing correction \cite{kst2,kt-02}.
Besides, a steep energy dependence of  
the dipole cross section $\sigma_{\bar qq}(r,s)$ (see Eqs.~(\ref{gbw-1})
and (\ref{kst-1})) especially at smaller dipole sizes $r$ causes
a steep energy rise of both corrections.

%
%%%%%%%%%%%%%%%%%%%%%%%%%%%%%%%%%%%%%%%%%%%%%%%%%%%%%%%%%%%%%%%%%%%%%%%%%%%
\section{Algorithm for numerical solution of the evolution
equation for the Green function}
\label{green}
%%%%%%%%%%%%%%%%%%%%%%%%%%%%%%%%%%%%%%%%%%%%%%%%%%%%%%%%%%%%%%%%%%%%%%%%%%%
%

As we mentioned in the previous section,
an explicit analytical expression for the Green function 
$G_{\bar qq}(\vec{r_2},z_2;\vec{r_1},z_1)$ 
(\ref{200})
can be found
only for the quadratic form (\ref{180}) of the dipole cross section
and for uniform nuclear density function (\ref{190}).
It was already analyzed in \cite{krt-98,krt-00} 
that such an approximation
should have a reasonable accuracy, especially for heavy nuclei.
We also discussed in the previous Section that
the higher accuracy can be achieved taking into account the fact
that the expression (\ref{230}) in the high energy limit can
be easily calculated using realistic parametrizations
of the dipole cross section 
(see Eqs.~(\ref{gbw-1}) and (\ref{kst-1})) 
and a realistic nuclear density function $\rho_A(b,z)$ \cite{saxon}.
Consequently, one needs to know the full Green function only
in the transition region from non-shadowing ($x_{Bj}\sim 0.1$)
to a fully developed shadowing given by Eq.~(\ref{230}).
Therefore the value of the energy dependent factor $C(s)$
in Eq.~(\ref{180}) was fixed \cite{krt-98,krt-00} 
separately for T and L photon polarizations (see Eq.~(\ref{224}))
in a such way that the asymptotic nuclear shadowing in DIS
is the same for the realistic parametrizations of the dipole
cross section Eqs.~(\ref{gbw-1}) and (\ref{kst-1}) 
and for approximation (\ref{180}).  
Correspondingly, the value $\rho_0$ of the uniform nuclear density 
(\ref{190}) was fixed in an analogical way as given by Eq.~(\ref{228})
and described shortly in the previous Section.
Such a procedure for determination of the factor $C(s)$ and $\rho_0$
was applied also in \cite{knst-01,n-02}
with respect to incoherent and coherent production
of vector mesons off nuclei.

In order to remove the above mentioned uncertainties
one should solve the evolution equation for the Green function
numerically for arbitrary parametrization of the dipole
cross section and for realistic nuclear density function.
However, the tax for this general solution is that one
does not obtain a nice analytical form for the Green function.
First we present an algorithm for the exact numerical
solution of the evolution equation. Using this algorithm we will 
calculate the nuclear shadowing in DIS and study
how the new results change in comparison with
predictions \cite{krt-98} based on above
mentioned approximations leading to harmonic oscillator 
Green function (\ref{200}).

In the process of numerical solution of the Schr\"odinger
equation (\ref{150}) for the Green function 
$G_{\bar qq}(\vec{r_2},z_2;\vec{r_1},z_1)$
it is not very convenient to treat 
the initial condition (\ref{155}) with 
two-dimensional Delta function on the r.h.s. 
In order to remove this problem one should use the
following substitutions
%
%%%%%%%%%%%%%%%%%%%%%%%%%%%%%%%%%%%%%%%%%%%%%%%%
\BA
g_1(\vec{r_2},z_2;z_1) = 
\int\,d^2r_1\,K_0(\epsilon\,r_1)\,\sigma_{\bar qq}(r_1,s)\,
G_{\bar qq}(\vec{r_2},z_2;\vec{r_1},z_1)\, ,
%----------
\label{310}
%----------
\EA
%%%%%%%%%%%%%%%%%%%%%%%%%%%%%%%%%%%%%%%%%%%%%%%%
%
and
%
%%%%%%%%%%%%%%%%%%%%%%%%%%%%%%%%%%%%%%%%%%%%%%%%
\BA
\frac{\vec{r_2}}{r_2}\,g_2(\vec{r_2},z_2;z_1) = 
\int\,d^2r_1\,K_1(\epsilon\,r_1)\,\sigma_{\bar qq}(r_1,s)\,
\frac{\vec{r_1}}{r_1}\,G_{\bar qq}(\vec{r_2},z_2;\vec{r_1},z_1)\, .
%----------
\label{320}
%----------
\EA
%%%%%%%%%%%%%%%%%%%%%%%%%%%%%%%%%%%%%%%%%%%%%%%%
%
Consequently, after some algebra with Eq.~(\ref{150})
one can introduce new 
functions $g_1(\vec{r_2},z_2;z_1)$
and $g_2(\vec{r_2},z_2;z_1)$ which satisfy now the following
evolution equations
%
%%%%%%%%%%%%%%%%%%%%%%%%%%%%%%%%%%%%%%%%%%%%%%%%%%%%%%%%%%%%%%%%%%%%%%
\BE
i\frac{d}{dz_2}\,g_1(\vec{r_2},z_2;z_1)=
\left\{\frac{1}{2\,\mu_{\bar qq}}
\left[\epsilon^{2} - 
\frac{\partial^2}{\partial\,r_2^2} 
- \frac{1}{r_2}\,\frac{\partial}{\partial\,r_2}\right]
+ V_{\bar qq}(z_2,\vec r_2,\alpha)\right\}
g_1(\vec{r_2},z_2;z_1)\ 
%----------
\label{330}
%----------  
 \EE
%%%%%%%%%%%%%%%%%%%%%%%%%%%%%%%%%%%%%%%%%%%%%%%%%%%%%%%%%%%%%%%%%%%%%%
%
and
%
%%%%%%%%%%%%%%%%%%%%%%%%%%%%%%%%%%%%%%%%%%%%%%%%%%%%%%%%%%%%%%%%%%%%%%
\BE
i\frac{d}{dz_2}\,g_2(\vec{r_2},z_2;z_1)=
\left\{\frac{1}{2\,\mu_{\bar qq}}
\left[\epsilon^{2} - 
\frac{\partial^2}{\partial\,r_2^2} 
- \frac{1}{r_2}\,\frac{\partial}{\partial\,r_2}
+ \frac{1}{r_2^2}\right]
+ V_{\bar qq}(z_2,\vec r_2,\alpha)\right\}
g_2(\vec{r_2},z_2;z_1)\ ,
%----------
\label{340}
%----------  
 \EE
%%%%%%%%%%%%%%%%%%%%%%%%%%%%%%%%%%%%%%%%%%%%%%%%%%%%%%%%%%%%%%%%%%%%%%
%
with the boundary conditions
%
%%%%%%%%%%%%%%%%%%%%%%%%%%%%%%%%
\beq
g_1(\vec{r_2},z_2;z_1)|_{z_2=z_1}=
K_0(\epsilon\,r_2)\,\sigma_{\bar qq}(r_2,s)
%----------
\label{350}
%----------
\eeq
%%%%%%%%%%%%%%%%%%%%%%%%%%%%%%%%
%
and
%
%%%%%%%%%%%%%%%%%%%%%%%%%%%%%%%%
\beq
g_2(\vec{r_2},z_2;z_1)|_{z_2=z_1}=
K_1(\epsilon\,r_2)\,\sigma_{\bar qq}(r_2,s)\, .
%----------
\label{360}
%----------
\eeq
%%%%%%%%%%%%%%%%%%%%%%%%%%%%%%%%
%
In Eqs.~(\ref{330}) and (\ref{340}) the quantity 
%
%%%%%%%%%%%%%%%%%%%
\beq
\mu_{\bar qq} = \nu\,\alpha\,(1-\alpha)
%----------
\label{365}
%----------
\eeq
%%%%%%%%%%%%%%%%%%%
%
plays the role of the reduced mass of the $\bar qq$ pair.
Consequently, the expression (\ref{260}) for total
photoabsorption cross section on a nucleus now reads
%
%%%%%%%%%%%%%%%%%%%%%%%%%%%%%%%%%%%%%%%%%%%%%%%%%%%%%%%
\BA
\sigma^{\gamma^*A}(x_{Bj},Q^2) &=&
A\,\sigma^{\gamma^*N}(x_{Bj},Q^2) - \Delta\,\sigma(x_{Bj},Q^2) 
\nonumber \\
&=& A\,\int\,d^2r\,\int_{0}^{1}\,d\alpha\,\sigma_{\bar qq}(r,s)
\,\Biggl (\Bigl |\Psi^T_{\bar qq}(\vec{r},\alpha,Q^2)\Bigr |^2 +
\Bigl |\Psi^L_{\bar qq}(\vec{r},\alpha,Q^2)\Bigr |^2\Biggr )
\nonumber \\
&-& \frac{3\,\alpha_{em}}{(2\pi)^2}\,\sum_{f=1}^{N_f}\,Z_f^2\,Re\,
\int\,d^2b\,\int_{-\infty}^{\infty}\,dz_1\,\int_{z_1}^{\infty}\,
dz_2\,\int_{0}^{1}\,d\alpha\,\int\,d^2r_2
\nonumber \\
&&\times\,\rho_A(b,z_1)\,\rho_A(b,z_2)\,\sigma_{\bar qq}(r_2,s)\,
\nonumber \\
&&\times\,\Biggl\{\Bigl[\,\alpha^2 + (1 - \alpha)^2\,\Bigr]
\,\epsilon^2\,
K_1(\epsilon\,r_2)\,g_2(\vec{r_2},z_2;z_1)
%----------
\label{370}
%----------
\\
&&\,\,\,\,\,\,\,\,\,\,\,\,\,\,\,
 + \,\Bigl[\,m_f^2 + 4\,Q^2\,\alpha^2\,(1 - \alpha)^2\,\Bigr]\,
K_0(\epsilon\,r_2)\,g_1(\vec{r_2},z_2;z_1)\Biggr\}\, .
\nonumber
\EA
%%%%%%%%%%%%%%%%%%%%%%%%%%%%%%%%%%%%%%%%%%%%%%%%%%%%%%%
%

There are several approaches for solving the time-dependent
one-dimensional
Schr\" odinger equation (see for example \cite{mc1,mc2,mc3,mc4}). 
One can not adopt directly these approaches for our purposes
because one needs to treat the time-dependent two-dimensional
Schr\" odinger equation (see Eq.~(\ref{150})).
Therefore, 
we will consider a modification of the method based on 
the Crank-Nicholson algorithm \cite{giordano}.
Details of this method for numerical solution of Eqs.~(\ref{330})
and (\ref{340}) are presented in the Appendix A.

%
%%%%%%%%%%%%%%%%%%%%%%%%%%%%%%%%%%%%%%%%%%%%%%%%%%%%%%%%%%
\section{Numerical results}
\label{results}
%%%%%%%%%%%%%%%%%%%%%%%%%%%%%%%%%%%%%%%%%%%%%%%%%%%%%%%%%%
%

As we mentioned above the main goal of this paper is
to present for the first time 
the realistic predictions for nuclear shadowing
based on exact numerical solutions of the
evolution equation for the Green function.
These predictions will be confronted with 
approximate results
obtained using harmonic oscillatory form of the
Green function (\ref{200}).
The proposed algorithm (see Appendix A)
for numerical solution of the
evolution equation for the Green function gives a possibility
to calculate nuclear shadowing for arbitrary LC potential
$V_{\bar qq}(z,\vec{r},\alpha)$ and nuclear density function.
It allows to perform an independent cross-check whether the results 
calculated using quadratic form of the imaginary part, 
$Im\,V_{\bar qq}(z,\vec{r},\alpha) = - 1/2\,\rho_A(b,z)\,C\,r^2$ 
($C \approx 3$), of the LC potential
and uniform density function $\rho_A(b,z) = \rho_0 = 0.16\fm^{-3}$ correspond
to predictions for nuclear shadowing taken from \cite{krt-98} calculated using
Green function of the form (\ref{200}).
Therefore, we calculate nuclear shadowing for calcium
and lead as was done in \cite{krt-98}.

In order to realize one-to-one comparison with the results from 
the paper \cite{krt-98} we made a several assumptions.
As was mentioned in Section~(\ref{lc}) we neglect the
real part of the LC potential $V_{\bar qq}(z_2,\vec{r_2},\alpha)$
in the Schr\"odinger equation (\ref{150})\footnote{Consequently,
the real part of $V_{\bar qq}(z_2,\vec{r_2},\alpha)$
is neglected as well in differential equations
(\ref{330}) and (\ref{340}).}
analyzing DIS at medium
and large values of $Q^2$. We neglect also the effects 
of nuclear antishadowing 
assuming that they are beyond the shadowing dynamics.
The corresponding values of Bjorken $x_{Bj}$ cover
medium and medium large values, $x_{Bj}\in (0.001,0.1)$.
For this reason we omit the effects of GS which
are important at small $x_{Bj}$ (large $\nu$). 
Although gluons can give some
small (not negligible) contribution to nuclear shadowing at
lower limit ($x_{Bj}\sim 0.001$) of investigated interval of
$x_{Bj}$, for simplicity
we do not include them in calculations
as was done also in the paper \cite{krt-98}.

We use an algorithm for numerical solution of the Schr\"odinger
equation for the Green function described in 
Section~(\ref{green}) and Appendix~A.
Numerical solution of Schr\"odinger equation allows us
to use realistic nuclear density function 
and realistic parametrizations
of the dipole cross section, $\sigma_{\bar qq}(r,s)$. 
These parametrizations naturally 
incorporate the energy ($x_{Bj}-$ ) dependence of 
$\sigma_{\bar qq}(r,s)$ which was not included so far
in calculations (see \cite{krt-98})\footnote{The 
energy dependence of dipole cross section
was included only via energy-dependent factor $C(s)$ in
approximation (\ref{180}). The factor $C(s)$ was determined by the procedure 
described shortly in Section~\ref{lc} and presented in details in 
\cite{krt-00,r-00,knst-01}.}.
We took the nuclear density function
in Woods-Saxon form \cite{saxon}.
The realistic calculations of nuclear shadowing were performed
at two different parametrizations of the dipole cross section,
GBW \cite{gbw} given by Eq.~(\ref{gbw-1}) and KST \cite{kst2}
given by Eq.~(\ref{kst-1}).

\begin{figure}[tbh]
\includegraphics{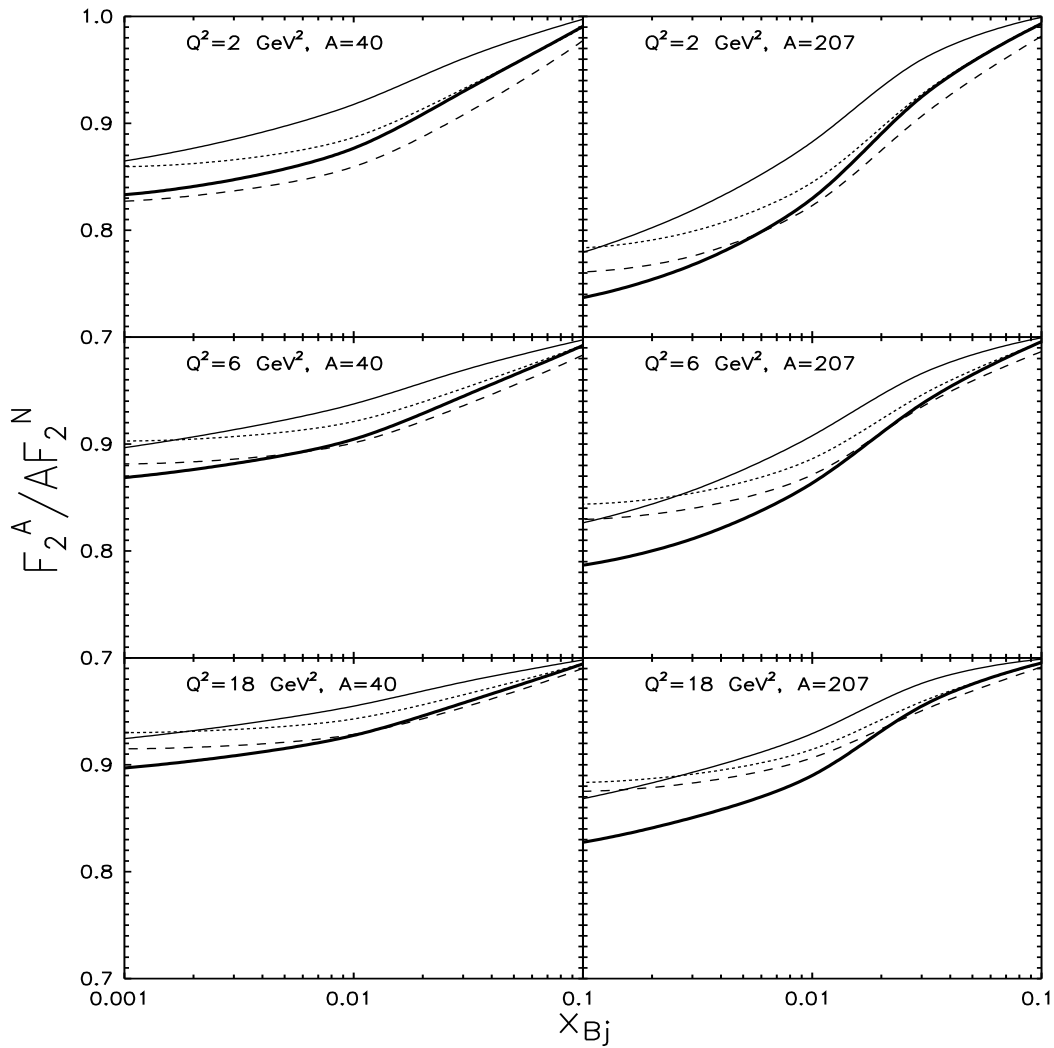}
\begin{center}
\vspace{14.0cm}
\parbox{13cm}
{\caption[Delta]
 {Nuclear shadowing for calcium and lead. The dashed curves
are calculated using harmonic oscillator Green function
approach (\ref{200}) corresponding to constant nuclear density
(\ref{190}) and dipole cross section (\ref{180})
with $C(s)\approx 3$ \cite{krt-98}.
The dotted curves are calculated for the same quadratic form
of the dipole cross section but for realistic nuclear density
function of the Woods-Saxon form \cite{saxon}. 
The thin and thick solid curves correspond to exact
numerical solution of the evolution equation for the Green function
using GBW \cite{gbw} and KST \cite{kst2} 
parametrization of the dipole cross section, respectively.
}
%%%%%%%%%%%%%%%%%%%%%%%%%
 \label{A-shad-ca-pb}}
%%%%%%%%%%%%%%%%%%%%%%%%%
\end{center}
 \end{figure}

Nuclear shadowing effects were studied via $x_{Bj}$-
behavior of the ratio of 
proton structure functions (\ref{270}) divided by the mass
number $A$.
The proton structure functions $F_2^N(x_{Bj},Q^2)$ and
$F_2^A(x_{Bj},Q^2)$ were calculated perturbatively
(we fixed the quark masses at $m_q=0.3\GeV$, $m_s=0.45\GeV$
and $m_c=1.5\GeV$)
via total photoabsorption cross sections
$\sigma^{\gamma^*N}(x_{Bj},Q^2)$ and 
$\sigma^{\gamma^*A}(x_{Bj},Q^2)$ given
by Eqs.~(\ref{120}) and (\ref{370}), respectively.
The results of calculations are shown in Fig.~{\ref{A-shad-ca-pb}}.
The dashed curves represent the predictions 
based on the harmonic oscillator Green
function (\ref{200}) approach corresponding 
to a constant nuclear density function
$\rho_0 = 0.16\,fm^{-3}$ and quadratic form of the dipole cross section (\ref{180})
with a constant factor $C(s)\approx 3$.
The dotted curves correspond to calculations using the same  
quadratic form of the dipole cross section but 
realistic nuclear density function of the Woods-Saxon form  
\cite{saxon}.
The thin and thick solid curve corresponds to realistic calculations
based on the exact numerical solution of the evolution
equation (\ref{150}) for the Green function using GBW
(\ref{gbw-1}) and KST (\ref{kst-1}) parametrization
of the dipole cross section, respectively.

%
%****************************************************************
%************************ FIG.2 *********************************
%****************************************************************
%****************************************************************
%

At low $x_{Bj}\lsim 0.001$ one should expect a saturation
of nuclear shadowing at level given by Eqs.~(\ref{230})
or (\ref{265}).
For parametrization (\ref{180}) of the dipole cross section
with constant $C(s)\approx 3$ \cite{krt-98},
this saturation level is fixed at some value depending
on $Q^2$ and the nuclear mass number $A$
(see the dashed and dotted lines in Fig.~\ref{A-shad-ca-pb}).   
However, it is not so for realistic parametrizations 
Eq.~(\ref{gbw-1}) and (\ref{kst-1})
where the saturation level is not fixed exactly due to
energy (Bjorken $x_{Bj}$-) dependence of the dipole cross section 
$\sigma_{\bar qq}(r,s)$. 

In the process of realistic calculations of nuclear shadowing
(\ref{370})
we tested the correctness of tremendous computations
based on the numerical evaluations of the functions
$g_1$ and $g_2$ from differential equations (\ref{330})
and (\ref{340}) in a such way that in the high energy
limit the results for nuclear shadowing must be the same 
as obtained from the expression (\ref{265}).

Results presented in Fig.~\ref{A-shad-ca-pb} show
quite a large deviation of the predictions within
harmonic oscillator Green function approach (dashed lines) from realistic
calculations performed for both parametrizations
of the dipole cross section (thin and thick solid lines).
This deviation depends on $Q^2$ and the nuclear mass
number $A$ as a result of quadratic form
of the dipole cross section 
(see Eq.~(\ref{180}) with $C(s)\approx 3$) 
and application of the
constant nuclear density function, $\rho_0=0.16\fm^{-3}$.
There is even not negligible difference between predictions
using constant and realistic form of the nuclear density function
(compare dashed and dotted lines). It allows to make conclusion
that the form of nuclear density function is also important
for model predictions. Note, that the dashed curves correspond
to predictions presented in the paper \cite{krt-98}. It is another
cross-check for correctness of calculations using above 
presented algorithm for numerical solution of the evolution equation
for the Green function.   

As one can see from Fig.~\ref{A-shad-ca-pb} 
at small and medium values of 
$Q^2$ ($Q^2 = 2$ and $6\GeV^2$), the approximate 
calculations
depicted by the dashed lines agree better with
realistic calculations using KST parametrization \cite{kst2}
of the dipole cross section expressed by Eq.~(\ref{kst-1}).
At large $Q^2 = 18\GeV^2$ and at 
$x_{Bj}\lsim 0.005$, however, the dashed lines 
seem to be in better agreement with realistic
calculations using GBW parametrization \cite{gbw}
given by Eq.~(\ref{gbw-1}). 
This fact confirms discussion presented in Section~\ref{lc}
that the GBW model is suited better at medium and large 
$Q^2\gsim 5\div 10\GeV^2$ 
and at medium small and small $x_{Bj}\lsim 0.01$
whereas the KST model prefers low and medium values of 
$Q^2\lsim 10\GeV^2$.

Nevertheless, calculations of nuclear shadowing 
using harmonic oscillator Green function
can be improved by determination of the energy-dependent
factor $C(s)$ in approximation (\ref{180}) by the procedure
mentioned shortly above in Sect.~\ref{lc} and described in details
in \cite{krt-00,r-00} for calculation of nuclear shadowing
in DIS and in \cite{knst-01,n-02} for calculation of
nuclear transparency for coherent and incoherent vector
meson production off nuclei.
That procedure allows to evaluate the factor $C(s)$ 
for each c.m. energy squared $s$ depending on
the values of $Q^2$ and $A$. As a result, at fixed $Q^2$
the parameter $C(s)$ rises
with $s$ as a consequence of energy- ($x_{Bj}$- ) dependent
realistic dipole cross section given by Eqs.~(\ref{gbw-1})
and (\ref{kst-1}). 
Thus, the value of $C(s)$
at $s$ corresponding to $x_{Bj}\in (0.001,0.01)$
exceeds so the fixed value $C(s)\approx 3$ used in predictions in 
\cite{krt-98}.
This fact should lead to a larger nuclear shadowing at $x_{Bj}\in 
(0.001,0.01)$
in comparison with what is shown 
in Fig.~\ref{A-shad-ca-pb} by the dashed lines.

As we mentioned above in Section~\ref{lc} the difference
between the thin and thick solid lines in Fig.~\ref{A-shad-ca-pb}
can be treated as a measure of the theoretical uncertainty
in the kinematic region where the both realistic parametrizations
GBW (\ref{gbw-1}) and KST (\ref{kst-1}) are applicable.
Therefore, it would be very useful for the future 
realistic calculations to connect advantages of both parametrizations
in the modified model for dipole cross section which can
be then safely used for all dipole sizes covering perturbative 
as well as nonperturbative region.

%
%%%%%%%%%%%%%%%%%%%%%%%%%%%%%%%%%%%%%%%%%%%%%%%%%%%%%%%%%%
\section{Summary and conclusions}
\label{conclusions}
%%%%%%%%%%%%%%%%%%%%%%%%%%%%%%%%%%%%%%%%%%%%%%%%%%%%%%%%%%
%

We present a rigorous quantum-mechanical approach based
on the light-cone QCD Green function formalism
which naturally incorporates the interference 
effects of CT and CL.
Within this approach \cite{krt-98,rtv-99,krt-00} 
we study nuclear shadowing in deep-inelastic scattering
at moderately small Bjorken $x_{Bj}$. 

Calculations of nuclear shadowing performed so far
were based only on the efforts to solve the evolution
equation for the Green function analytically.
Analytical harmonic oscillatory form of the Green
function (\ref{200}) could be obtained only taking into account
additional approximations like a constant nuclear
density function (\ref{190}) and the dipole cross
section of the quadratic form (\ref{180}). 
It brings additional 
theoretical uncertainties in predictions
for nuclear shadowing. 
In order to remove these uncertainties 
we solve the evolution equation for the Green
function numerically.

We perform for the first time the exact numerical
solution of the evolution equation
for the Green function using two realistic
parametrizations of the dipole cross section 
(GBW \cite{gbw} and KST \cite{kst2}) 
and realistic nuclear density function 
of the Woods-Saxon form \cite{saxon}.
This exact numerical solution 
does not require to put any additional approximations.
Analyzing only
medium and large values of $Q^2$
we neglect the real part of the LC potential 
$V_{\bar qq}(z_2,\vec{r_2},\alpha)$ in the 
time-dependent two-dimensional
Schr\"odinger equation (\ref{150})
responsible for interaction between $\bar q$ and $q$.
We neglect also the nuclear antishadowing effect
as was done in \cite{krt-98} assuming that it
is beyond the shadowing dynamics.
Performing calculations at medium and medium large
values of $x_{Bj}\in (0.001,0.1)$
we neglect for simplicity also the contribution
of the higher Fock states leading to effects of GS.
This is supported also by the one-to-one 
comparison of the realistic calculations
with the predictions from the paper 
\cite{krt-98},
where GS is neglected as well.

In order to compare the realistic calculations
with data on nuclear shadowing, the effects of
GS should be taken into account
especially at $x_{Bj}\lsim 0.001$.
The same path integral technique 
\cite{kst2} can be applied in this case.
However, the calculations of 
GS (see \cite{krtj-02},
for example) were performed so far
using analogical approximations as already 
mentioned above like
a constant nuclear density function and
the quadratic form (see Eq.~(\ref{180}))
of the dipole gluon-gluon-nucleon 
cross section, $\sigma_{gg}(r,s) = 9/4\,\sigma_{\bar qq}(r,s)$ plus
further assumptions which simplify the
final expression for GS.
Moreover, the GS was calculated from the shadowing
of the $|\bar qqG\ra$ Fock component of a longitudinally
polarized photon at sufficiently large $Q^2$
where the three-body Green function, $G_{\bar qqG}$, 
is assumed to be factorized
as a product of two-body ones \cite{kst2}.
Using the algorithm presented above one can calculate
GS exactly for the general case of nuclear shadowing
for a three-parton system, i.e. one can solve numerically
the Schr\"odinger equation for the Green function
$G_{\bar qqG}$ describing propagation of the $\bar qqG$ system
through a nuclear medium.
We are going to calculate numerically the gluon contribution
to nuclear shadowing in a forthcoming paper.

We present analogical 
numerical results of nuclear shadowing in DIS 
with correct quantum-mechanical treatment 
of multiple interaction of the virtual photon
fluctuations and of the nuclear form factor
as was done in \cite{krt-98,krt-00}.
We found quite large differences (see Fig.~\ref{A-shad-ca-pb})
between realistic predictions 
and the approximate results 
obtained within harmonic
oscillator Green function approach (\ref{200}).
At small and medium values of 
$Q^2$ the approximate predictions
agree better with realistic calculations 
using KST parametrization \cite{kst2}
of the dipole cross section Eq.~(\ref{kst-1}).
At large $Q^2$, however, they
seem to be in better agreement with realistic
calculations using GBW parametrization \cite{gbw}
given by Eq.~(\ref{gbw-1}). 
It confirms the fact 
that the GBW model is well suited at medium and large 
$Q^2$ and at medium small and small $x_{Bj}$,
whereas the KST model prefers low and medium values of 
$Q^2$.
Therefore, the future realistic calculations 
require to revise existing parametrizations
for dipole cross section in order to be 
used for whole region of dipole sizes. 
 
Concluding, the universality of the LC dipole approach
based on the Green function formalism allows us to apply
the presented algorithm 
for the exact numerical solution
of the evolution equation for the Green function 
also for calculations of other processes like Drell-Yan production,
vector meson production etc. including the effects of gluon
shadowing at high energies as well.

\medskip

\noindent
 {\bf Acknowledgments}:
We are grateful to Alexander Tarasov 
for stimulating discussions.
This work has
been supported in part by the Slovak Funding Agency, Grant No. 2/2099/22 and
Grant No. 2/1169/21.

%%%%%%%%%%%%%%%%%%%%%%%%%%%%%%%%%%%%%%%%%%%%%%%%%%%%%%%%%%
 \def\appendix{\par
 \setcounter{section}{0} \setcounter{subsection}{0}
 \def\thesection{Appendix \Alph{section}}
\def\thesubsection{\Alph{section}.\arabic{subsection}}
\def\theequation{\Alph{section}.\arabic{equation}}
\setcounter{equation}{0}}
%%%%%%%%%%%%%%%%%%%%%%%%%%%%%%%%%%%%%%%%%%%%%%%%%%%%%%%%%%

 \appendix

\section{Description of the method for numerical
solution of the time-dependent Schr\"odinger equation}

We treat here only differential equation (\ref{330})
for the function $g_1(\vec{r_2},z_2;z_1)$ and  
describe in details the method for its numerical solution.
This method is then analogically applicable  
for numerical solution of Eq.~(\ref{340})
for the function $g_2(\vec{r_2},z_2;z_1)$.

Looking at Eq.~(\ref{330}),
one needs to solve numerically the following
time-dependent Schr\"odinger equation
\footnote{We put $r_2\equiv r$ and
$z_2$ plays the role of time $t$.}
%
%%%%%%%%%%%%%%%%%%%%%%%%
\beq
i\,\frac{d}{dt}\,g_1(\vec{r},t) = \hat{H}\,g_1(\vec{r},t)\, ,
%----------
\label{a10}
%---------
\eeq
%%%%%%%%%%%%%%%%%%%%%%%%
%
where $\hat{H}$ is the Hamiltonian operator defined by
%
%%%%%%%%%%%%%%%%%%%%%%%%
\beq
\hat{H} = \frac{1}{2\,\mu_{\bar qq}}\,\Biggl [\epsilon^2 -
\frac{\partial^2}{\partial r^2} - \frac{1}{r}\,\frac{\partial}
{\partial r}\,\Biggr ] + V_{\bar qq}(r,t)\, .
%----------
\label{a20}
%---------
\eeq
%%%%%%%%%%%%%%%%%%%%%%%%
%
Here the complex LC potential $V_{\bar qq}(r,t)$ is assumed to have only 
the imaginary part responsible for absorption of $\bar qq$ photon
fluctuation in the nuclear medium
(see discussion in Section~\ref{lc} and Eq.~(\ref{170})),
%
%%%%%%%%%%%%%%%%%%%%%%%%%%%%%%
\beq
V_{\bar qq}(r,t) = - \frac{i}{2}\,\sigma_{\bar qq}(r,s)\,\rho_A(b,t)\, ,
%----------
\label{a30}
%---------
\eeq
%%%%%%%%%%%%%%%%%%%%%%%%
%
where $b$ is the nuclear impact parameter and $r$ is the transverse
separation between $\bar q$ and $q$ at the point $z_2$.
The longitudinal coordinate $z_2$
plays the role of time $t$ for the $\bar qq$ pair propagation 
from the point $z_1$.
In Eq.~(\ref{a20}) the quantity $\mu_{\bar qq}$ is expressed by 
Eq.~(\ref{365}).

Ignoring for a moment the fact that $\hat{H}$ is an operator,
Eq.~(\ref{a10}) has the formal solution,
%
%%%%%%%%%%%%%%%%%%%%%%%%%%%%%%
\beq
g_1(\vec r,t) = exp\,( - i\,t\,\hat{H})\,g_1(\vec{r},0)\, ,
%----------
\label{a40} 
%----------
\eeq
%%%%%%%%%%%%%%%%%%%%%%%%%%%%%%
%
where $g_1(\vec{r},0)$ is the function at $t=0$.
Thus, if one knows $g_1(\vec r,0)$, one can formally
calculate the behavior at all future times using
Eq.~(\ref{a40}). Unfortunately, this formal solution
is not of much practical use since the Taylor
expansion of the exponential factor in Eq.~(\ref{a40})
involves a very large (infinite) number of terms.
However, it does suggest a way to proceed numerically.
Let us consider a formal solution applying over a very small time
interval. After time discretization in steps $\Delta t$
using Eq.~(\ref{a40}) we obtain
%
%%%%%%%%%%%%%%%%%%%%%%%%%%%%%%%%%%%%%
\beq
g_1(\vec r,t+\Delta t) = exp\,( - i\,\Delta t\,\hat{H})\,
g_1(\vec{r},t)\, .
%----------
\label{a50}
%----------
\eeq
%%%%%%%%%%%%%%%%%%%%%%%%%%%%%%%%%%%%%
%
Consequently, at sufficiently small time intervals $\Delta t$
the higher order terms in Taylor expansion of the
exponential factor in (\ref{a40}) are small enough 
and can be neglected. Then we can include only the term
linear in $\hat{H}$,
%
%%%%%%%%%%%%%%%%%%%%%%%%%%
\beq
exp\,( - i\,\Delta t\,\hat{H}) \approx 1 - i\,\Delta t\,\hat{H}\, .
%----------
\label{a60}
%----------
\eeq
%%%%%%%%%%%%%%%%%%%%%%%%%%
%
However, this way of approximating the exponential factor
is not correct with regard to maintaining unitarity \cite{giordano}.
In order to settle this problem one should use an approach
that satisfy unitarity writing the exponential
factor in Eq.~(\ref{a50}) in what is known as the Cayley form
%
%%%%%%%%%%%%%%%%%%%%%%%%%%%%%%
\beq
exp\,( - i\,\Delta t\,\hat{H}) \approx 
\frac{1 - \frac{1}{2}\,i\,\Delta t\,\hat{H}}
{1 + \frac{1}{2}\,i\,\Delta t\,\hat{H}}\, .
%----------
\label{a70}
%----------
\eeq
%%%%%%%%%%%%%%%%%%%%%%%%%%%%%%
%
Using this approximation and Eq.~(\ref{a40}) 
to propagate the function $g_1(\vec{r},t)$
forward in time we obtain
%
%%%%%%%%%%%%%%%%%%%%%%%%%%%%%%
\beq
g_1(\vec{r},t+\Delta t) \approx 
\frac{1 - \frac{1}{2}\,i\,\Delta t\,\hat{H}}
{1 + \frac{1}{2}\,i\,\Delta t\,\hat{H}}\,
g_1(\vec{r},t)\, .
%----------
\label{a80}
%----------
\eeq
%%%%%%%%%%%%%%%%%%%%%%%%%%%%%%
%
This expression will be the basis for numerical approach.
From (\ref{a80}) we first obtain
%
%%%%%%%%%%%%%%%%%%%%%%%%%%%%%%%%%%%%%
\beq
\Biggl [1 + \frac{1}{2}\,i\,\Delta t\,\hat{H}\,\Biggr ]\,
g_1(\vec{r},t+\Delta t) = 
\Biggl [1 - \frac{1}{2}\,i\,\Delta t\,\hat{H}\,\Biggr ]\,
g_1(\vec{r},t)\, .
%----------
\label{a90}
%----------
\eeq
%%%%%%%%%%%%%%%%%%%%%%%%%%%%%%
%
 
Given (\ref{a90}),
a natural way to proceed is to discretize also space into units
of size $\Delta r$ and write the function $g_1$ as
$g_1(r,t)\equiv g_1(m\,\Delta r,n\,\Delta t)$.
Then one can express the first and second derivatives included
in the Hamiltonian operator (\ref{a20}) in the usual
finite-difference form :
%
%%%%%%%%%%%%%%%%%%%%%%%%%%%%%%%%%%%%
\BA
\frac{\partial}{\partial\,r}\,g_1(r,t) \approx
\frac{g_1(r+\Delta r) - g_1(r,t)}{\Delta r}
= 
\frac{g_1(m+1,n) - g_1(m,n)}{\Delta r}
%-----------
\label{a100}
%-----------
\EA
%%%%%%%%%%%%%%%%%%%%%%%%%%%%%%%%%%%%
%
and
%
%%%%%%%%%%%%%%%%%%%%%%%%%%%%%%%%%%%%
\BA
\frac{\partial^2}{\partial\,r^2}\,g_1(r,t) &\approx&
\frac{\partial}{\partial\,r}\,\Biggl [
\frac{g_1(r+\Delta r) - g_1(r,t)}{\Delta r}\,\Biggr ] =
\frac{g_1(r+\Delta r,t) - 2\,g_1(r,t) + g_1(r-\Delta r,t)}{(\Delta 
r)^2}
\nonumber \\
&=&
\frac{g_1(m+1,n) - 2\,g_1(m,n) + g_1(m-1,n)}{(\Delta r)^2}
%-----------
\label{a110}
%-----------
\EA
%%%%%%%%%%%%%%%%%%%%%%%%%%%%%%%%%%%%
%

If one replaces in Eq.~(\ref{a90}) the Hamiltonian operator by
(\ref{a20}), converting everything to finite-deference form
using also Eqs.~(\ref{a100}) and (\ref{a110}), and
rearranging a few terms one obtains the following expression
%
%%%%%%%%%%%%%%%%%%%%%%%%%%%%%%%%%%%%%%%%%%%%
\BA
g_1(m+1,n+1) &+& h(m)\,\Biggl [\,2\,i\,\lambda -
2\,\mu_{\bar qq}\,(\Delta r)^2\,V(m,n+1) - \epsilon^2\,(\Delta r)^2
- 2 - \frac{1}{m}\,\Biggr ]\,
g_1(m,n+1) 
\nonumber \\
&+& h(m)\,g_1(m-1,n+1)
\nonumber \\
= - g_1(m+1,n) &+& h(m)\,\Biggl [\,2\,i\,\lambda +
2\,\mu_{\bar qq}\,(\Delta r)^2\,V(m,n) + \epsilon^2\,(\Delta r)^2
+ 2 + \frac{1}{m}\,\Biggr ]\,
g_1(m,n) 
\nonumber \\
&-& h(m)\,g_1(m-1,n)\, ,
%-----------
\label{a120}
%-----------
\EA
%%%%%%%%%%%%%%%%%%%%%%%%%%%%%%%%%%%%%%%%%%%%
%
where the function $h(m)$ is given by
%
%%%%%%%%%%%%%%%%%%%%%%%
\beq
h(m) = \frac{m}{1 + m}
%-----------
\label{a130}
%-----------
\eeq
%%%%%%%%%%%%%%%%%%%%%%%
%
and
%
%%%%%%%%%%%%%%%%%%%%%%%
\beq
\lambda = \frac{2\,(\Delta r)^2\,\mu_{\bar qq}}{\Delta t}
%-----------
\label{a140}
%-----------
\eeq
%%%%%%%%%%%%%%%%%%%%%%%
%
with the reduced mass of $\bar qq$ pair defined by Eq.~(\ref{365}).

The algorithm very effective for solving the time-dependent
Schr\"odinger equation in one dimension
is known as the Crank-Nicholson method described
in details in \cite{giordano,gss-67}. 
However, in order to solve (\ref{a120}) one should modify this method
for more complicated case of two dimensional Schr\"odinger
equation. 
We begin by defining a shorthand for the r.h.s. of Eq.~(\ref{a120})
%
%%%%%%%%%%%%%%%%%%%%%%%%%%%%%%%%%%%%%%%%%%%%
\BA
\Omega(m,n) &\equiv&
- g_1(m+1,n) 
+ h(m)\,\Biggl [\,2\,i\,\lambda +
2\,\mu_{\bar qq}\,(\Delta r)^2\,V(m,n) + \epsilon^2\,(\Delta r)^2
+ 2 + \frac{1}{m}\,\Biggr ]\,
g_1(m,n) 
\nonumber \\
&-& h(m)\,g_1(m-1,n)\, ,
%-----------
\label{a150}
%-----------
\EA
%%%%%%%%%%%%%%%%%%%%%%%%%%%%%%%%%%%%%%%%%%%%
%
in order to rewrite Eq.~(\ref{a120}) as
%
%%%%%%%%%%%%%%%%%%%%%%%%%%%%%%%%%%%%%%%%%%%%
\BA
g_1(m+1,n+1) &+& h(m)\,\Biggl [\,2\,i\,\lambda -
2\,\mu_{\bar qq}\,(\Delta r)^2\,V(m,n+1) - \epsilon^2\,(\Delta r)^2
- 2 - \frac{1}{m}\,\Biggr ]\,
g_1(m,n+1) 
\nonumber \\
&+& h(m)\,g_1(m-1,n+1)\,\, = \,\, \Omega(m,n)\, .
%-----------
\label{a160}
%-----------
\EA
%%%%%%%%%%%%%%%%%%%%%%%%%%%%%%%%%%%%%%%%%%%%
%

For convenient numerical procedure one should
write $g_1(m+1,n+1)$ as a function
of just $g_1(m,n+1)$ in the following form
%
%%%%%%%%%%%%%%%%%%%%%%%%%%%%%%%%%%%%%%
\beq
g_1(m+1,n+1) = e(m,n)\,g_1(m,n+1) + f(m,n)\, ,
%-----------
\label{a170}
%-----------
\eeq
%%%%%%%%%%%%%%%%%%%%%%%%%%%%%%%%%%%%%%
%
and so one can calculate $g_1(m+1,n+1)$ directly
from $g_1(m,n+1)$.

If one inserts Eq.~(\ref{a170}) into Eq.~(\ref{a160})
and does a little arithmetic, one can find that the
factors $e(m,n)$ and $f(m,n)$ must be given by
the following implicit relations
%
%%%%%%%%%%%%%%%%%%%%%%%%%%%%%%%%%%%%%%
\beq
e(m,n) = h(m)\,\Biggl [2 + \frac{1}{m} +
2\,\mu_{\bar qq}\,(\Delta r)^2\,V(m,n+1) + \epsilon^2\,(\Delta r)^2
- 2\,i\,\lambda
- \frac{1}{e(m-1,n)}\Biggr ]
%-----------
\label{a180}
%-----------
\eeq
%%%%%%%%%%%%%%%%%%%%%%%%%%%%%%%%%%%%%%
%
and
%
%%%%%%%%%%%%%%%%%%%%%%%%%%%%%%%%%%%%%%
\beq
f(m,n) = 
\Omega(m,n) +
h(m)\,\frac{f(m-1,n)}{e(m-1,n)} \, .
%-----------
\label{a190}
%-----------
\eeq
%%%%%%%%%%%%%%%%%%%%%%%%%%%%%%%%%%%%%%
%

One supposes that a propagation of $\bar qq$ pair
in the nuclear medium is confined to some region
of space so that spatial index runs from $m=0$ to
$m=M$ and imposes the boundary conditions
$g_1(0,n) = g_1(M,n) = 0$.
The expressions (\ref{a180}) and (\ref{a190})
for the factors $e(m,n)$ and $f(m,n)$ can be applied
only in the interior of the system.
Consequently, from the boundary condition for the function
$g_1$ at $m=0$ together with Eqs.~(\ref{a150}) and
(\ref{a170}) one can find that at this end of the
system
%
%%%%%%%%%%%%%%%%%%%%%%%%%%%%%%%%%%
\beq
e(1,n) = 
h(1)\,\Biggl [3 +
2\,\mu_{\bar qq}\,(\Delta r)^2\,V(1,n+1) + \epsilon^2\,(\Delta r)^2
- 2\,i\,\lambda\,\Biggr ]
%-----------
\label{a200}
%-----------
\eeq
%%%%%%%%%%%%%%%%%%%%%%%%%%%%%%%%%%%%%%
%
and
%
%%%%%%%%%%%%%%%%%%%%%%%%%%%%%%%%%%%%%%
\beq
f(1,n) = 
\Omega(1,n)\, .
%-----------
\label{a210}
%-----------
\eeq
%%%%%%%%%%%%%%%%%%%%%%%%%%%%%%%%%%%%%%
%

For the first time step, $n=0$, the factors $\Omega(1,0)$,
$e(1,0)$ and $f(1,0)$ can be explicitly calculated from
the initial function, $g_1(m,0) = \sigma_{\bar qq}(m\,\Delta r,s)\,
K_0(\epsilon\,m\,\Delta r)$, which is assumed to be given
as an initial condition (see also Eq.~(\ref{350})).
Using known values of $e(1,0)$ and $f(1,0)$ 
one can calculate then the factors 
$e(2,0)$ and $f(2,0)$ 
from the implicit expressions 
Eqs.~(\ref{a180}) and (\ref{a190}) and continue so
for all values of $m$ along the system.
Hence, we traverse the system from $m=0$ to $m=M$,
to calculate $e(m,0)$ and $f(m,0)$ for all $m$.

For the further purposes, Eq.~(\ref{a170}) can
be rearranged in the following form
%
%%%%%%%%%%%%%%%%%%%%%%%%%%%%%%%
\beq
g_1(m,n+1) = \frac{g_1(m+1,n+1) - f(m,n)}{e(m,n)}\, .
%-----------
\label{a220}
%-----------
\eeq
%%%%%%%%%%%%%%%%%%%%%%%%%%%%%%%
%
As was mentioned above, at the end of the system $m=M$ the function 
$g_1$ vanishes. Consequently, one can write
%
%%%%%%%%%%%%%%%%%%%%%%%%%%%%%%%%%%%%%%
\beq
g_1(M-1,n+1) = \frac{g_1(M,n+1) - f(M-1,n)}{e(M-1,n)} = 
- \,\frac{f(M-1,n)}{e(M-1,n)}\, ,
%-----------
\label{a230}
%-----------
\eeq
%%%%%%%%%%%%%%%%%%%%%%%%%%%%%%%
%
since $g_1(M,n) = 0$ for all values of $n$.
One can thus use Eq.~(\ref{a230}) to obtain $g_1(M-1,1)$,
which is the value of the new function one spatial unit in
from the "right" boundary, $m=M$.
Then one can use Eq.~(\ref{a220}) to calculate the function $g_1$
at $m=M-2,M-3$, etc., as one traverses the system backward from large
to small values of $m$.

\begin{figure}[tbh]
\includegraphics{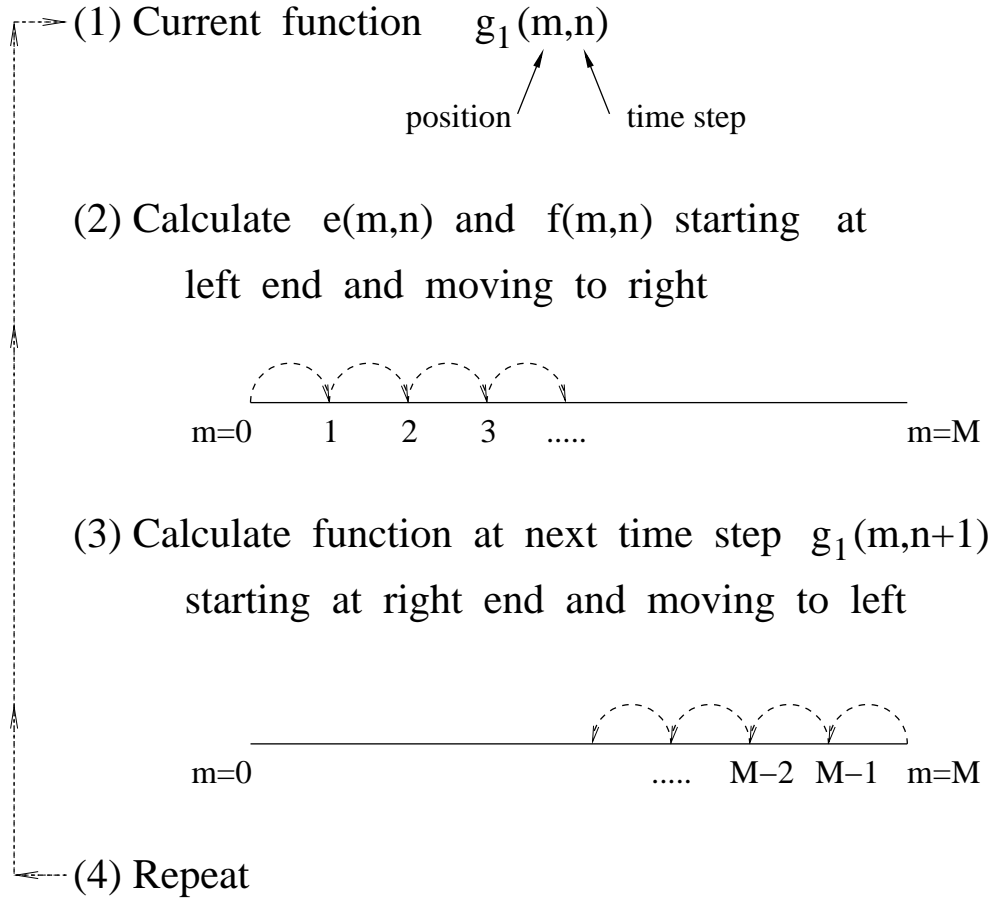}
\begin{center}
\vspace{12.3cm}
\parbox{13cm}
{\caption[Delta]
 {Schematic description of the Cranck-Nicholson algorithm.
}
%%%%%%%%%%%%%%%%%%%%%%%%%
 \label{algorithm}}
%%%%%%%%%%%%%%%%%%%%%%%%%
\end{center}
 \end{figure}

Finally, the algorithm can be summarized in the following way
(see also the schematic description in Fig.~{\ref{algorithm}}):

\begin{enumerate}

\item
One begins with the initial function $g_1(m,0)$ given by Eq.~(\ref{350}).

\item
The system is traversed from small to large values of $m$ and
the functions $e(m,0)$ and $f(m,0)$ are calculated using 
Eqs.~(\ref{a200}) and (\ref{a210}) initially and
Eqs.~(\ref{a180}) and (\ref{a190}) thereafter.

\item
The system is traversed from large to small values of $m$
and $g_1(m,1)$ is calculated using Eq.~(\ref{a230}) initially
and Eq.~(\ref{a220}) thereafter.
This completes one iteration and yields the function $g_1$ at
$n=1$ ($t=\Delta t$).

\item
Steps (2) and (3) are repeated to obtain the function $g_1$
as a function of time ($n\geq 1$).

\end{enumerate}

%
%****************************************************************
%************************ FIG.3 *********************************
%****************************************************************
%****************************************************************
%

\end{document}